\documentclass[aps,prd,showpacs,letterpaper,amsmath,
twocolumn]{revtex4}
\usepackage{graphicx}

\begin{document}

\title{Study of highly-excited string states at the Large Hadron Collider}

\author{Douglas M. Gingrich}

\altaffiliation[Also at ]{TRIUMF, Vancouver, BC V6T 2A3 Canada} 

\email{gingrich@ualberta.ca}

\author{Kevin Martell}

\affiliation{Centre for Particle Physics, Department of Physics, University
of Alberta, Edmonton, AB T6G 2G7 Canada}

\date{\today}

\begin{abstract}
In TeV-scale gravity scenarios with large extra dimensions, black holes
may be produced at future colliders.
Good arguments have been made for why general relativistic black holes
may be just out of reach of the Large Hadron Collider (LHC). 
However, in weakly-coupled string theory, highly excited string states
-- string balls -- could be produced at the LHC with high rates and
decay thermally, not unlike general relativistic black holes.
In this paper, we simulate and study string ball production and decay at
the LHC.
We specifically emphasize the experimentally-detectable similarities and
differences between string balls and general relativistic black holes at
a TeV scale.
\end{abstract}


%\keywords{string balls, black holes, string theory, extra dimensions,
%beyond Standard Model}

\maketitle

%%%%%%%%%%%%%%%%%%%%%%%%%%%%%%%%%%%%%%%%%%%%%%%%%%%%%%%%%%%%%%%%%%%%%%%%%%%%%%%
\section{Introduction\label{sec1}}

Models of large-extra dimensions~\cite{Arkani98,Antoniadis98,Arkani99}
and experimental constraints~\cite{Kapner,ALEPH,DELPHI,L304,CDF,D0} 
allow the scale of gravity to be as low as a TeV. 
At this scale, the intriguing possibility exists that black holes could
be produced at the Large Hadron Collider (LHC) and in cosmic-ray 
events~\cite{Argyres,Banks,Dimopoulos01,Giddings01a,Feng01,Anchordoqui01}. 
This possibility has been well studied for the case in which the black
hole is treated semiclassically, and is produced and decays according
to the concepts of general relativity (GR).
In this paper, we consider black holes as GR objects, and we will
address the issue of what happens below the energy scale at which
trans-Planckian objects are no longer considered GR black holes.  

Meade and Randall~\cite{Meade} have recently provided a nice summary of
the conditions required on the masses of gravitational objects for them
to be considered GR black holes. 
One commonly used condition is that the entropy of the black hole 
must be greater than 25 in order to fulfill the thermodynamic
description of black holes~\footnote{For Schwarzschild black holes, the
entropy requirement also ensures that the Compton wavelength of the
black hole is less than its horizon radius.}.  
This leads to the black hole mass production requirement of
$M_\mathrm{BH} \gtrsim 5 M_D$, where $M_D$ is the fundamental Planck
scale in higher dimensions.   
If $M_D$ is about 1~TeV, imposing the GR condition on black holes leads
to a requirement on the black hole mass of $M_\mathrm{BH} \gtrsim
5$~TeV. 
Thus, because of this requirement, high mass GR black holes may not be
accessible to the LHC. 
Their production at the LHC will be particularly unlikely if black holes
are not produced through totally inelastic collisions and some of the
parton energy is not available for black hole
formation~\cite{Anchordoqui03,Anchordoqui04,Gingrich06a}.    

Due to the steeply falling parton density distributions in the protons
with increasing parton centre of mass energy, the black hole cross
section drops with increasing black hole mass.
This means that the most probable black holes produced are those just
satisfying the GR condition and thus closest to the fundamental Planck
scale. 
Unfortunately, the most accessible black holes are also the least
theoretically understood.

We will define a mass threshold at which black holes can no longer be
treated by GR.
Below this GR threshold we enter the regime of quantum gravity.
In this regime, Meade and Randall~\cite{Meade} have considered
black-hole type objects within composite models and have studied the
possibility of their decay into dijets (and di-leptons).
Another exciting possibility for this regime occurs when everything
below the GR threshold is treated in the context of weakly-coupled
string theory.  
Although string theory is not required in TeV-scale gravity in higher
dimensions, it does allow us to postulate how a black hole makes a
transition across the GR threshold as it evaporates.

The scenario of large extra dimensions is not a model but rather a paradigm
in which models can be built.
Although inspired by string theory, the large extra dimensions
paradigm is not based on it.
However, embedding the large extra dimensions into string theory could
provide an understanding of the strong-gravity regime and a picture of
the evolution of a black hole at the last stages of
evaporation~\cite{Bowick86}. 
In this picture, black holes end their Hawking evaporation when their
mass reaches a critical mass.
At this point they transform into high-entropy string states -- string
balls -- without ever reaching the singular zero-mass limit.   

It has been shown that string states produced below the GR threshold
could have a cross section comparable to that of the black
hole~\cite{Dimopoulos02,Cheung02p,Chamblin02}. 
Hence, these states will be even more accessible than black holes at the 
LHC. 
Moreover, even if black holes are produced at the LHC, they will evolve
into these string states.
In themselves, string balls are interesting because they are a new form
of matter involving gravity and string theory.
However, if large extra dimensions are realized in a string theory of
quantum gravity, excited string states of Standard Model particles will
have TeV masses~\cite{Cullen,Burikham,
Anchordoqui07,
Anchordoqui08a,
Anchordoqui08c}. 
These states could provide the first signatures of low-scale quantum
gravity.  

Cheung~\cite{Cheung02p}, and Chamblin and Nayak~\cite{Chamblin02}
calculated large string ball cross sections at the LHC based on the
parton cross sections of Dimopoulos and
Emparan~\cite{Dimopoulos02}~\footnote{String ball production has also
been discussed in the context of ultra-high energy neutrinos in cosmic
radiation~\cite{Anchordoqui05}.}.   
However, full simulations were not performed, and experimental and
background effects were not discussed; that is the goal of this paper. 
With the startup of the LHC such detailed studies are timely and of
value.

This paper is structured as follows.
In Sec.~\ref{sec2}, we summarize the weakly-coupled string model of highly
excited string states embedded in large extra dimensions.
The hierarchy of the energy scales involved are discussed in Sec.~\ref{sec3}. 
In Sec.~\ref{sec4}, string ball production is described, while in
Sec.~\ref{sec5} a model for string ball evaporation is developed.
Our results are presented in Sec.~\ref{sec6} and a discussion follows in 
Sec.~\ref{sec7}. 
We do not discuss excited string resonances of Standard Model particles,
nor string effects near or below the string scale in this paper. 

%%%%%%%%%%%%%%%%%%%%%%%%%%%%%%%%%%%%%%%%%%%%%%%%%%%%%%%%%%%%%%%%%%%%%%%%%%%%%%%
\section{Weakly-Coupled Strings\label{sec2}}

If the energy of parton-parton scattering is comparable to the string
scale, the point-particle description of scattering will have to be
replaced by a string-string description of scattering.
As the energy increases above the string scale, the string states will
become highly excited, jagged, and entangled.
Such string states are commonly referred to as string
balls~\cite{Dimopoulos02}.  
Eventually at high enough energies a transition point is reached in
which the string ball turns into a black hole.

Embedding TeV-scale gravity scenarios in realistic string models could
enable calculations near $M_D$.
One perturbative string theory with weak-scale string tension is the SO(32)
type-I theory having 

\begin{enumerate}
\item new dimensions much larger than the weak scale,
\item $\mathcal{O}(1)$ string coupling,
\item Standard Model fields identified with open strings localized on a
      3-brane, and  
\item a gravitational sector consisting of closed strings propagating
      freely in the extra dimensions.
\end{enumerate}

\noindent
At present, there are no $\mathcal{N} = 1$ supersymmetric string models in
accordance with the large extra dimensions scenario, that break to only
the Standard Model at low energies without the presence of extra
massless particles. 
However, there are non-supersymmetric string models that can realize the
large extra dimensions scenario and break to only the Standard Model at
low energy with no extra massless matter~\cite{Kokorelis04,Cremades02}.

We will consider a model with $n$ large extra dimensions and $(6-n)$ small
extra dimensions.
After compactification of the small dimensions to the size of the string 
length scale, we obtain a relationship between the fundamental Planck
scale and the string parameters:

\begin{equation} \label{eq1}
M_D^{n+2} \sim \frac{M_\mathrm{s}^{n+2}}{g_\mathrm{s}^2}\, ,
\end{equation}

\noindent
where $n$ is the number of large extra dimensions, $M_\mathrm{s}$ is the
string scale, and $g_\mathrm{s}$ is the string coupling.
The string coupling constant is determined by the expectation value of
the dilaton field.
If string theory is perturbative ($g_\mathrm{s} < 1$), we see that
$M_\mathrm{s} < M_D$ for all $n$. 
Eq.~(\ref{eq1}) is an equality to order unity.
The exact numerical coefficient, which may depend on $n$, is model
dependent.
It depends on the string theory and compactification scheme, and would
involve one-loop calculations.  
The value of the numerical coefficient normally does not matter since $M_D$
and $M_\mathrm{s}$ are not well defined masses but are energy scales at
which new phenomena occur.  
However, for the numerical calculations performed in this paper it is
important to take into account all the $\mathcal{O}(1)$ numerical
factors in a consistent manner.  
If the coefficient is greater than unity, then $M_D$ is greater than
$M_\mathrm{s}$.   
If the coefficient is less than unity, there are values of
$g_\mathrm{s}$ for which $M_\mathrm{s}$ is greater than $M_D$.
As in previous work, we shall take the coefficient to be unity, and thus
$M_D \ge M_\mathrm{s}$.
This assumption breaks down only for the case of $g_\mathrm{s}$ greater
than the coefficient, at which point the \lq\lq highly-excited
string\rq\rq\ states are actually black hole states.

%%%%%%%%%%%%%%%%%%%%%%%%%%%%%%%%%%%%%%%%%%%%%%%%%%%%%%%%%%%%%%%%%%%%%%%%%%%%%%%
\subsection{Correspondence Principle}

According to the string theory of quantum gravity, the minimum mass
above which a black hole can be treated general relativistically
is~\cite{Susskin93,Horowitz96}   

\begin{equation} \label{eq2}
M_\mathrm{min} \sim \frac{M_\mathrm{s}}{g_\mathrm{s}^2}\, .
\end{equation}

\noindent
The properties of a black hole with mass $M_\mathrm{min}$ matches those
of a string ball with the same mass.
This is called the correspondence principle and the mass at which this
happens is the correspondence point.
When a black hole makes a transition to a string it can become a single
string, multiple strings, or radiation.
The single string configuration dominates since its entropy is the
highest~\cite{Horowitz96,Dimopoulos02}. 

The number of microstates of both black holes and string balls should be
the same at the correspondence point.
The entropy of a long string is proportional to its mass:

\begin{equation} \label{eq3}
S_\mathrm{s} \sim \sqrt{\alpha^\prime} M = \frac{M}{M_\mathrm{s}}\, , 
\end{equation}

\noindent
where $\alpha^\prime$ is the slope parameter given in terms of the
string tension $T$ as $\alpha^\prime = 1/(2\pi T)$.
We have used $\sqrt{\alpha^\prime} = \ell_\mathrm{s} = 1/M_\mathrm{s}$.  
At the correspondence point $S_\mathrm{s} \sim 1/g_\mathrm{s}^2$.

The Bekenstein entropy~\cite{Bekenstein73} of a black hole is
proportional to its area.
The black hole entropy in higher dimensions is 

\begin{eqnarray} \label{eq4}
S_\mathrm{BH} & = & \frac{4\pi}{n+2} f(n) \left( \frac{M}{M_D}
\right)^\frac{n+2}{n+1}\nonumber\\
& \sim & \frac{4\pi}{n+2} f(n) \frac{1}{g_\mathrm{s}^2} \left(
\frac{g_\mathrm{s}^2 M}{M_\mathrm{s}} \right)^\frac{n+2}{n+1}\, ,
\end{eqnarray}

\noindent
where

\begin{equation} \label{eq5}
f(n) \equiv \left[ \frac{2^n \pi^\frac{n-3}{2} \Gamma \left( \frac{n+3}{2}
\right)}{n+2}  \right]^\frac{1}{n+1} 
\end{equation}

\noindent
is an $n$-dependent factor of $\mathcal{O}(1)$.
In the last expression of Eq.~(\ref{eq4}), we have used the relationship 
between the fundamental Planck scale and the string parameters given by 
Eq.~(\ref{eq1}).
At the correspondence point $S_\mathrm{BH} \sim 1/g_\mathrm{s}^2$ and is
equal to the string entropy at the correspondence point to within a
numerical coefficient.   
The black hole and string entropies have the same mass dependence
near the correspondence point but the numerical coefficient is unknown. 
A better understanding of the string state near the string scale would
be required to precisely determine the coefficient.

The coefficient multiplying the correspondence point in Eq.~(\ref{eq2})
is usually considered to be a factor of order unity.
The factor represents exactly when the string state forms a black hole.
The factor depends on the string theory in terms of its entropy and the
relationship between the string scale and the fundamental
higher-dimensional Planck scale. 
For a factor of unity, the black hole entropy is always smaller than the
string entropy at the correspondence point for common superstring
theories.  
We will take the coefficient in Eq.~(\ref{eq2}) to be unity as in
previous studies.   
If this assumption is invalid and the coefficient is much less than
unity, the energy range in which the effects of quantum gravity are
important will be very narrow. 
If the coefficient is much larger than unity, black holes are unlikely to
be observed at the LHC even if the Planck and string scales are around a
TeV. 

Although the correspondence principle has worked well in four
dimensions when numerical coefficients have been dropped, it has not
worked so well for a consistent set of coefficients in higher
dimensions.  
Halyo \textit{et al}.~\cite{Halyo} could not get an exact match between
the four dimensional Schwarzschild or non-extreme charged black holes
and any fundamental string theory. 
Solodukhin~\cite{Solodukhin98} obtained matching by adding a
logarithmic quantum correction to the black hole entropy.
A gravitation term was then added to the string entropy to get exact 
matching of the two terms in the black hole entropy.
Adding the extra degrees of freedom allowed matching.
This matching only works in four dimensions where the self-interactions
are an integer power series in $g_\mathrm{s}$.

Presumably we must match the black hole entropy to the interacting
string entropy rather than the free entropy.
It is natural to identify the ``internal states'' of the black hole not
with states of the free string but with a part of the states of the
interacting string.
We can consider the string entropy as a perturbation series with respect 
to $g_\mathrm{s}$. 
In principle, the string entropy could possess some non-perturbative
corrections, behaving as $\sim 1/g_\mathrm{s}^2$. 

It is also possible that the black hole states can only turn into a
subset of the available string states.
If the subset is large enough, the entropies should only differ by a
numerical coefficient of order unity.
Thus a one-to-one correspondence between black hole states and string
states my not be necessary.

%%%%%%%%%%%%%%%%%%%%%%%%%%%%%%%%%%%%%%%%%%%%%%%%%%%%%%%%%%%%%%%%%%%%%%%%%%%%%%%
\subsection{Random Walk String}

The black hole's size at the correspondence point is of the order of the
string length scale $\ell_\mathrm{s}$.
By contrast, long excited string states are a chain of connected string
bits, each with length equal to  $\ell_\mathrm{s}$.
An excited string has a tendency to spread out as in a random walk.
Since one end of each string bit will perform a random walk relative to
the other, the size of these objects must be calculated statistically.
The step size of the random walk is $\ell_\mathrm{s}$ and its total
length is $(M/M_\mathrm{s})\ell_\mathrm{s}$, so the mean radius of the
average configuration of mass $M$ (i.e.\ the radius of the string ball)
is~\cite{Horowitz98,Damour99}  

\begin{equation} \label{eq6}
R_\mathrm{rw} \sim \sqrt{\frac{M}{M_\mathrm{s}}}\,
\ell_\mathrm{s} \sim \sqrt{M}\, \ell_\mathrm{s}^{\,3/2} >
\ell_\mathrm{s}\, .
\end{equation}

\noindent
However, this neglects the gravitational self-interaction of the string.
This is responsible for keeping the string compact at a size of about
$\ell_\mathrm{s}$ near the correspondence point.
Hence, shortly after the black hole to string ball transition, the
string abruptly ``puffs-up'' from string length scale size $\ell_\mathrm{s}$
to random-walk size $R_\mathrm{rw}$. 

At the LHC, the puff-up might not be too large or happen at all.
That is, the black hole horizon radius at the correspondence point is
$f(n)\ell_\mathrm{s}$, where $f(n) = 1.3-2.4$ for $n=3-6$, while the
maximum random-walk string size is about $\ell_\mathrm{s}/g_\mathrm{s}$.
Thus for small values of $g_\mathrm{s}$, the puff-up would not occur. 

%%%%%%%%%%%%%%%%%%%%%%%%%%%%%%%%%%%%%%%%%%%%%%%%%%%%%%%%%%%%%%%%%%%%%%%%%%%%%%%
\section{Energy Scales\label{sec3}}

The production of black holes and string states depends on four free
parameters $M_\mathrm{s}, g_\mathrm{s}, n$, and $M_D$. 
Typically 

\begin{equation} \label{eq7}
M_\mathrm{s} 
< M_D 
< \frac{M_\mathrm{s}}{g_\mathrm{s}} 
< \frac{M_\mathrm{s}}{g_\mathrm{s}^2}\, . 
\end{equation}

\begin{figure}[htb]
\begin{center}
%\setlength{\unitlength}{0.96pt}
%\setlength{\unitlength}{0.53pt}
%\begin{picture}(460,55)(-20,-30)
%\thicklines
%\put(-20,-30){\line( 1, 0){460}}
%\put(440,-30){\line( 0, 1){ 55}}
%\put(440, 25){\line(-1, 0){460}}
%\put(-20, 25){\line( 0,-1){ 55}}
%\put(0,0){\line(1,0){420}}
%\put(  0,-5){\line(0,1){10}}
%\put( 30,-5){\line(0,1){10}}
%\put( 60,-5){\line(0,1){10}}
%\put( 90,-5){\line(0,1){10}}
%\put(270,-5){\line(0,1){10}}
%\put(300,-5){\line(0,1){10}}
%\put(420,-5){\line(0,1){10}}
%\put(  0,10){\makebox(0,0)[b]{0 TeV}}
%\put(300,10){\makebox(0,0)[b]{10 TeV}}
%\put(420,10){\makebox(0,0)[b]{14 TeV}}
%\put( 30,-10){\makebox(0,0)[t]{$M_\mathrm{s}$}}
%\put( 60,-10){\makebox(0,0)[t]{$M_D$}}
%\put( 90,-10){\makebox(0,0)[t]{$\frac{M_\mathrm{s}}{g_\mathrm{s}}$}}
%\put(270,-10){\makebox(0,0)[t]{$\frac{M_\mathrm{s}}{g_\mathrm{s}^2}$}}
%\end{picture}
\includegraphics[width=\columnwidth]{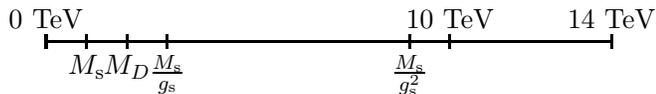}
\caption{Energy scales for black holes and string balls in the context
of the LHC energy.
Not shown are the compactification scale or the ultraviolet cutoff scale.}
\label{figscale}
\end{center}
\end{figure}

\noindent
Later on, we will see that $M_\mathrm{s}/g_\mathrm{s}$ divides the regions
between perturbative string theory and unitarity.
Figure~\ref{figscale} displays a possible hierarchy of energy scales at
the LHC. 

For GR black holes, we require their mass $M_\mathrm{BH} \ge \zeta M_D$,
and thus using Eq.~(\ref{eq2})

\begin{equation} \label{eq8}
M_\mathrm{min} = \frac{M_\mathrm{s}}{g_\mathrm{s}^2} = \zeta M_D\, .
\end{equation}

\noindent
Normally we will take $\zeta \sim 5$ as required by the thermodynamic
and Compton wavelength arguments given in Sec.~\ref{sec1}.
However, $\zeta \sim 3.4$ is acceptable for higher dimensions. 
If the unit coefficient in Eq.~(\ref{eq2}) is invalid, it can somewhat be
compensated for by the choice of $\zeta$.
If the assumed $\zeta$ is too low, black hole production will be
overestimated and string ball production will be slightly underestimated
in this paper.

It is not known how much higher above $M_\mathrm{s}$ the energy must be
in order to be in the stringy regime, where random-walk string states
are valid.  
As $M/M_\mathrm{s}$ drops to unity this picture will no longer be valid.
We take

\begin{equation} \label{eq9}
M_\mathrm{SB} > 3 M_\mathrm{s}
\end{equation}

\noindent
as a requirement for the validity of the long and jagged picture of
string balls.
The exact location of this low-energy cutoff mainly affects integrated
cross sections.
Most of the results in this paper are insensitive to this choice of
cutoff provided it is not too large, in which case neither string balls
nor black holes will be observable at the LHC.

If string theory is strongly coupled ($g_\mathrm{s} \sim 1$), all the
scales would be about the same. 
In this case, black hole like behavior will appear at the scale
$M_D$ and all the string states will be black hole states. 
For very weak string coupling ($g_\mathrm{s} \ll 1$), string ball
production occurs at very high energies and there would be no black hole  
production at the LHC. 
For moderate values of $g_\mathrm{s}$ there is a significant energy range
between $M_\mathrm{s}$ and $M_\mathrm{min}$ inside which the spectrum is
intrinsically stringy and the GR approximation fails.
The fundamental Planck scale $M_D$ is typically smaller than
$M_\mathrm{min}$ so the GR approximation can fail even above $M_D$.
We will consider $g_\mathrm{s}$ moderately less than unity and thus a
significant stringy regime between the scales $M_\mathrm{s}$ and
$M_\mathrm{min}$ can occur. 

Virtual graviton effects depend on the ultraviolet cutoff of the
Kaluza-Klein (KK) spectrum. 
Since the fundamental Planck scale in models of large extra dimensions
is $M_D \sim 1$~TeV, it is natural to expect this cutoff to be of the
same order. 
It is possible that the ultraviolet cutoff is somewhat lower than
$M_D$. 
We will take an operational approach of considering $M_\mathrm{s}$ to be
the lowest energy scale at which the first effects of gravity would be
observed in experiments. 
This approach requires $M_\mathrm{s}$ to be consistent with experimental
limits on gravitational effects and thus $M_\mathrm{s} \gtrsim 1$~TeV. 

Using the two constraints (Eq.~(\ref{eq1}) and Eq.~(\ref{eq8})) between 
$M_\mathrm{s}$, $g_\mathrm{s}$, $n$, and $M_D$, the string ball
characteristics depend on two independent parameters; we normally choose 
$M_\mathrm{s}$ and $n$.  
Although $M_D$ is the more common independent scale, by using
$M_\mathrm{s}$ we are working with the lowest energy scale.
All higher energy scales will be determined by the constraints as well
as the string coupling.  
The string coupling and Planck scale are determined by

\begin{equation} \label{eq10}
g_\mathrm{s}^2 = 1/\zeta^\frac{n+2}{n+1}
\end{equation}

\noindent
and

\begin{equation} \label{eq11}
M_D = \zeta^\frac{1}{n+1} M_\mathrm{s}\, .
\end{equation}

\noindent
The choice of $\zeta=5$ keeps the values for $g_\mathrm{s}$ in the
perturbative regime and $M_D$ consistent with experimental limits.
Table~\ref{tab1} shows values for the parameters $g_\mathrm{s}$, $M_D$, 
$M_\mathrm{s}/g_\mathrm{s}$, and $M_\mathrm{s}/g_\mathrm{s}^2$
when $M_\mathrm{s} = 1$~TeV and $\zeta = 5$.  
The significance of $M_\mathrm{s}/g_\mathrm{s}$ will be explained in the
next section.
Also shown in Table~\ref{tab1} are values of $f(n)$ given by
Eq.~(\ref{eq5}), which are often ignored. 

\begin{table}[htb]
\caption{\label{tab1}Parameter values for $M_\mathrm{s} = 1$~TeV and
$\zeta = 5$.} 
\begin{ruledtabular}
\begin{tabular}{cccccc}
$n$ & $g_\mathrm{s}$ & $M_D$ [TeV] & $M_\mathrm{s}/g_\mathrm{s}$ [TeV] &
$M_\mathrm{s}/g_\mathrm{s}^2$ [TeV] & $f$\\ \hline
%0 & 0.20 & 5.0 & 5.0 & 25.0 & 0.07958\\
%1 & 0.30 & 2.2 & 3.3 & 11.2 & 0.46066\\ \hline
2 & 0.34 & 1.7 & 2.9 &  8.6 & 0.90856\\
3 & 0.37 & 1.5 & 2.7 &  7.5 & 1.33746\\
4 & 0.38 & 1.4 & 2.6 &  6.9 & 1.73470\\
5 & 0.39 & 1.3 & 2.6 &  6.5 & 2.10164\\
6 & 0.40 & 1.3 & 2.5 &  6.3 & 2.44219\\
\end{tabular}
\end{ruledtabular}
\end{table}

%%%%%%%%%%%%%%%%%%%%%%%%%%%%%%%%%%%%%%%%%%%%%%%%%%%%%%%%%%%%%%%%%%%%%%%%%%%%%%%
\section{String Ball Production\label{sec4}}

The correspondence principle also suggests that the production cross
section for string balls will match the black hole cross section at
centre of mass energies around $M_\mathrm{s}/g_\mathrm{s}^2$:

\begin{equation} \label{eq12}
\sigma(\mathrm{SB})|_{M_\mathrm{SB}=M_\mathrm{s}/g_\mathrm{s}^2} \sim
\sigma(\mathrm{BH})|_{M_\mathrm{BH}=M_\mathrm{s}/g_\mathrm{s}^2}\, .
\end{equation}

\noindent
Because the black hole size near the correspondence point is smaller
than the excited string size, the transition may involve the effects
of strong self-gravity around this energy~\cite{Damour99}.    

The production cross section for string balls with mass between the
string scale $M_\mathrm{s}$ and $M_\mathrm{s}/g_\mathrm{s}$ grows
with the centre of mass energy squared $\hat{s} = M^2$
as~\cite{Dimopoulos02} 

\begin{equation} \label{eq13}
\sigma_\mathrm{s} \sim \frac{g_s^2}{M_\mathrm{s}^4} M^2\, .
\end{equation}

\noindent
This expression generalizes to arbitrary dimensions as a consequence
of the independence of the string scattering amplitude on dimensions.

Most scenarios of low-scale quantum gravity as low-energy effective
theories are valid only up to order $M_\mathrm{s}$.
Above this scale, the naive calculations typically violate
perturbative unitarity.
The unitarity bound for Eq.~(\ref{eq13}) occurs at
$g_\mathrm{s}^2\hat{s}/M_\mathrm{s}^2$~\cite{Amati87}. 
Thus the production cross section for string balls grows with $M^2$
only for $M_\mathrm{s} < M \le M_\mathrm{s}/g_\mathrm{s}$.
One has to introduce some ad hoc unitarization scheme, since a
fundamental string theory is still unavailable.  
At the unitarity point, $M_\mathrm{s}/g_\mathrm{s}$, the string cross
section is 

\begin{equation} \label{eq14}
\sigma_\mathrm{s} = \frac{1}{M_\mathrm{s}^2}\, .  
\end{equation}

\noindent
We thus take the string ball cross section to be constant between
$M_\mathrm{s}/g_\mathrm{s}$ and $M_\mathrm{s}/g_\mathrm{s}^2$. 
The proton-proton cross section for string ball production is dominated
by the parton density functions at low production masses.
Thus if the assumptions in Eq.~(\ref{eq14}) are incorrect and the cross
section continues to rise slowly, this effect will be hard to detect in
experimental data.
The exact coefficient (normalization) is a question to be determined by
experiment. 

The approximate match between this constant cross section and that of a
black hole at the correspondence point becomes obvious when using the
relationship between $M_D$ and $M_\mathrm{s}$ given by Eq.~(\ref{eq1})
in the black hole cross section: 

\begin{eqnarray}\label{eq15}
\sigma_\mathrm{BH} & = & \pi \frac{f^2(n)}{M_D^2} \left( \frac{M}{M_D}
\right)^\frac{2}{n+1}  
\to \pi \frac{f^2(n)}{M_\mathrm{s}^2} 
\left( \frac{g_\mathrm{s}^2 M}{M_\mathrm{s}} \right)^\frac{2}{n+1}\nonumber\\
& \sim & l_\mathrm{s}^2 \left(
\frac{g_\mathrm{s}^2 M}{M_\mathrm{s}} \right)^\frac{2}{n+1}\, ,
\end{eqnarray}

\noindent
where we have dropped numerical coefficients in the last expression.

Including the $n$-dependent coefficient for the black hole cross section,
the parton cross section over all three energy regions is

\begin{eqnarray} \label{eq16}
\hat{\sigma} = \left\{
\begin{array}{ll}
\frac{g_\mathrm{s}^2 M^2}{M_\mathrm{s}^4} & 
M_\mathrm{s} \ll M \le 
\frac{M_\mathrm{s}}{g_\mathrm{s}}\, ,\\ 
\frac{1}{M_\mathrm{s}^2} & 
\frac{M_\mathrm{s}}{g_\mathrm{s}} \le M \le
\frac{M_\mathrm{s}}{g_\mathrm{s}^2}\, ,\\ 
\pi \frac{f^2(n)}{M_D^2} \left( \frac{M}{M_D}
\right)^\frac{2}{n+1} & 
\frac{M_\mathrm{s}}{g_\mathrm{s}^2} < M\, .\\
\end{array}
\right.
\end{eqnarray}

\noindent
The lower two energy ranges lead to string ball production and the
higher one leads to black hole production.

Dimopoulos and Emparan~\cite{Dimopoulos02} considered the cross sections
to within only numerical factors.
This allowed them to match the cross sections at the correspondence
point but changed the standard black hole cross section. 
Cheung~\cite{Cheung02p} multiplied the string cross sections by the
$n$-dependent numerical factor $f^2(n)$ to obtain matching.
This resulted in string cross sections that depended on the number of
large extra dimensions in an arbitrary way. 

The cross section should contain Chan-Paton factors~\cite{Paton}, which
control the projection of the initial state onto the string
spectrum~\cite{Burikham}. 
In general, this projection is not uniquely determined by the
low-lying particle spectrum.
This is usually accounted for by introducing one or more arbitrary
constants.
Besides $g_\mathrm{s}$ and $M_\mathrm{s}$, the Veneziano amplitudes are 
characterized by two constants which parameterize the Chan-Paton traces
for string models. 
Thus the overall coefficient in the cross section, Eq.~(\ref{eq13}), is
unknown and we will take it to be unity.
The exact value of the cross section is an experimental issue.

The saturation cross section is $1/M_\mathrm{s}^2$.
If exact continuity of the cross section is required, the
correspondence point will occur at very low energy.
For all $n$ values, the correspondence point will be so low that it
rules out string ball production at the LHC.
A discontinuity in the cross section is not physical but with a better
knowledge of the strong gravitational effects, we could expect the
transition to be smooth. 

%%%%%%%%%%%%%%%%%%%%%%%%%%%%%%%%%%%%%%%%%%%%%%%%%%%%%%%%%%%%%%%%%%%%%%%%%%%%%%%
\section{String Ball Evaporation\label{sec5}}

Highly-excited long strings (averaged over degenerate states of the same
mass) emit massless (as well as massive) particles with a thermal
spectrum at the Hagedorn temperature~\cite{Amati99}. 
Hence, the conventional description of evaporation in terms of blackbody
emission can be applied to highly excited string states.
Assuming a type-I string theory, the emissions can take place either in
the bulk (into closed strings) or on the brane (into open strings).

The Hagedorn temperature is given by

\begin{equation} \label{eq17}
T_\mathrm{s} = \frac{M_\mathrm{s}}{\sqrt{8}\pi}\, .
\end{equation}

\noindent
The temperature is the same for evaporation to open strings on the brane or
closed strings in the bulk.
The Hagedorn temperature can be viewed as either the maximum temperature
or the temperature of a phase transition.
However, one should be aware that the concept of temperature and phase
transition are ill-defined in the presence of gravity.

To leading order, the formal temperature, given by $T = (\partial
S/\partial M)^{-1}$, is usually equal to the Hagedorn temperature.
The formal temperature for a free string includes a mass dependent term
coming from a logarithmic term in the entropy.
This term raises the temperature for small masses.
The entropy could also include self-gravity and fixed-size terms.
Since strings emit particles with a thermal spectrum at the Hagedorn
temperature and not the formal temperature, we will ignore any extra
terms in the temperature.

The Hawking temperature for black hole evaporation in higher
dimensions is

\begin{eqnarray} \label{eq18}
T_\mathrm{H} & = & \frac{n+1}{4\pi} \frac{1}{R_\mathrm{h}} =
\frac{n+1}{4\pi f(n)} \left( \frac{M_D}{M} \right)^{n+1} M_D\nonumber\\
& \to &
\frac{n+1}{4\pi f(n)} \left( \frac{1}{g_\mathrm{s}^2} \right)^n
M_\mathrm{s}\, ,   
\end{eqnarray}

\noindent
where the last expression is the maximum temperature at the
correspondence point.
Similarly to the entropy and cross section, the Hagedorn temperature of
an excited string matches the Hawking temperature of a black hole at
the correspondence point. 
In the random-walk phase, the string still evaporates at the Hagedorn
temperature, which eventually brings the size down towards
$\ell_\mathrm{s}$.
Even if black holes are produced at the LHC, the black holes will
decay into string balls, and eventually down to low-lying string states. 

The Hagedorn temperature in Eq.~(\ref{eq17}) is for a free string.
Perturbative corrections to that temperature could be expected.
For a highly-excited, weakly-coupled, neutral string, the leading order
term in perturbation theory is just the Hagedorn temperature at which a 
very weakly-coupled string will radiate.
Thus the temperature of a string at a large level number would have a
perturbation expansion of the form

\begin{equation} \label{eq19}
T = T_\mathrm{s} + g_\mathrm{s}^2 F(M) + \cdots\, .
\end{equation}

The black hole decays mainly on the brane~\cite{Emparan00}.
The emissivity depends not only on the temperature (and thus mass) but
also weakly on the total number of space-time dimensions.  
An equal amount of radiation will be emitted into the KK tower of a
single bulk species as will be emitted into a single brane species.   
Including the number of degrees of freedom at the LHC, the bulk
emission is notable but still not
dominant~\cite{Cardoso06a,Gingrich07a}. 

The decay depends not only on the temperature but also on the
$D$-dimensional area of the object emitting Hawking radiation.
For black holes and random-walk strings, the area is well defined.
Moreover, because there is no relationship between the physical size of 
the string and its temperature, to first order, the decay may be
different from that of a black hole.
To examine this, we follow the arguments of Emparan, Horowitz, and
Myers~\cite{Emparan00} but apply them to random-walk strings.
In $D$-dimensions, the energy radiated by a blackbody of temperature
$T$ and surface area $A_D$ is 

\begin{equation} \label{eq20}
\frac{dE_D}{dt} = \sigma_D A_D T^D\, ,
\end{equation}

\noindent
where $\sigma_D$ is the $D$-dimensional Stefan-Boltzmann constant given
by

\begin{equation} \label{eq21}
\sigma_D = \frac{\Omega_{D-3}}{(2\pi)^{D-1}(D-2)} \Gamma(D) \zeta(D)\, ,
\end{equation}

\noindent
where $\zeta(D)$ is the Riemann zeta function, $\Omega_D$ denotes the
volume of a unit $D$-sphere, and $A_D = r^{D-2} \Omega_{D-2}$ is the
area of the string of radius $r$. 

\begin{figure}[htb]
\begin{center}
\includegraphics[width=\columnwidth]{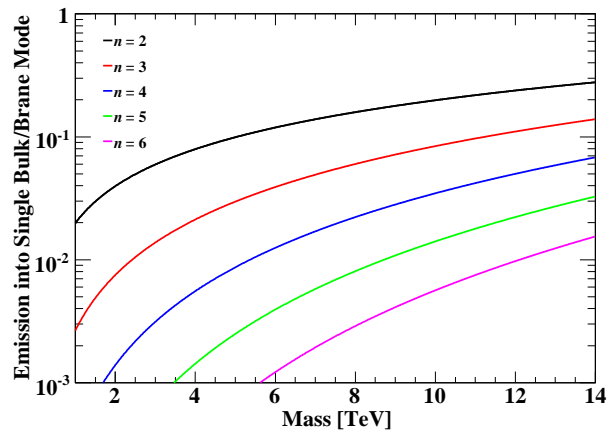}
\caption{Ratio of emission into a single bulk mode to a single brane mode
for highly excited strings.
The ratio decreases with increasing $n$.}
\label{figemiss}
\end{center}
\end{figure}

The ratio of emissivities in $D$-dimensions to four dimensions is
given by 

\begin{equation} \label{eq22}
\frac{\dot{E}_D}{\dot{E}_4} \sim \left( \frac{1}{\sqrt{8}\pi} \right)^n
\left( \frac{M}{M_\mathrm{s}} \right)^\frac{n}{2}\, ,
\end{equation}

\noindent
where the omitted numerical coefficient increases from 1 to about 3 as
$n$ increases from 0 to 6.
Shown in Fig.~\ref{figemiss} is the ratio of emissivities.
We see that the emissivity ratio is always less than unity at the
LHC. 
Only when $M \gtrsim 56 M_\mathrm{s}$ will the string radiate more into
a single bulk mode; this corresponds to $g_\mathrm{s} < 0.13$.

Just below the correspondence point, the string size is a maximum and
the ratio of emissivities becomes

\begin{equation} \label{eq23}
\left[ \frac{\dot{E}_D}{\dot{E}_4} \right]_\mathrm{max} \sim \left(
\frac{1}{\sqrt{8}\pi g_\mathrm{s}} \right)^n\, .
\end{equation}

\noindent
Thus $g_\mathrm{s} \lesssim 1/(\sqrt{8}\pi) = 0.1$ for the bulk modes to
dominate. 
This result was mentioned in Ref.~\cite{Dimopoulos02}.
The result is different from Fairbairn's~\cite{Fairbairn05} who
ignored all $n$-dependent coefficients. 

%%%%%%%%%%%%%%%%%%%%%%%%%%%%%%%%%%%%%%%%%%%%%%%%%%%%%%%%%%%%%%%%%%%%%%%%%%%%%%%
\section{Results\label{sec6}}

In this section, we present the results of a study of string balls at
the LHC based on the previously described model. 
The black hole and string ball parton cross sections were given in
Eq.~(16).
However, only a fraction of the total centre of mass energy $\sqrt{s}$ in
a proton-proton collision is available in the parton scattering
process. 
The total cross section can be obtained by convoluting the parton-level
cross section with the parton distribution functions, integrating over
the phase space, and summing over the parton types.
In this paper, we assume all the available parton energy $\sqrt{\hat{s}}$
goes into forming the black hole or string ball.
Although this might be unlikely, it avoids confusing the effects from
totally inelastic string ball production with unknown inelastic effects. 
Also, throughout this paper proton-proton collisions at 14~TeV centre of
mass energy are considered.
When referring to string ball production it is usually understood to
also include black hole production if the initial parton energy
is high enough. 

Throughout this paper we use the CTEQ6L1 (leading order with leading
order $\alpha_\mathrm{s}$) parton distribution functions~\cite{Pumplin}
within the LHAPDF framework~\cite{LHAPDF}.
The momentum scale for the parton distribution functions is set equal to
the black hole or string ball mass for convenience.
The extrapolation of the parton distribution functions into the
trans-Planckian or \lq\lq trans-stringian\rq\rq\ region based on Standard
Model evolution from present energies is questionable, since the
evolution equations neglect gravity.

\begin{figure}[htb]
\begin{center}
\includegraphics[width=\columnwidth]{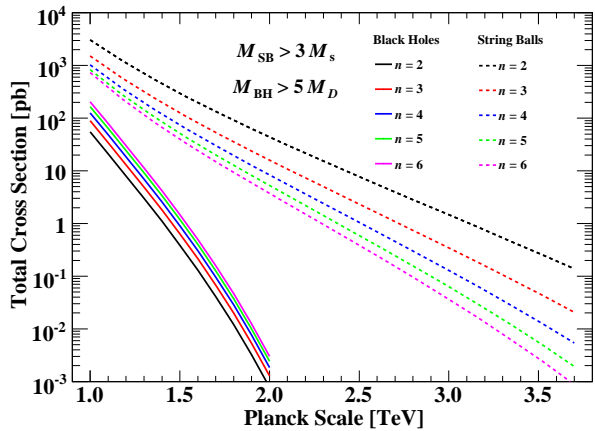}
\caption{Total proton-proton cross section versus Planck scale for the
production of black holes (solid curves) and string balls plus black
holes (dashed curves) for various numbers of extra dimensions $n$.
The black hole cross section increases with increasing $n$. 
The string ball cross section decreases with increasing $n$.
\label{figppMD}}
\end{center}
\end{figure}

Figure~\ref{figppMD} shows the total proton-proton cross section versus
Planck scale for the production of black holes and string balls for
various numbers of extra dimensions $n$. 
We see that the string ball plus black hole cross sections are at least an
order of magnitude higher, and that a substantially enhanced range of
$M_D$ could be probed with string balls at the LHC. 
The black hole cross section is weakly dependent on $n$, while the $n$
dependence of the string ball cross section is mainly due to the
$n$-dependent relationship between the string scale and the Planck scale
Eq.~(\ref{eq11}).  

\begin{figure}[htb]
\begin{center}
\includegraphics[width=\columnwidth]{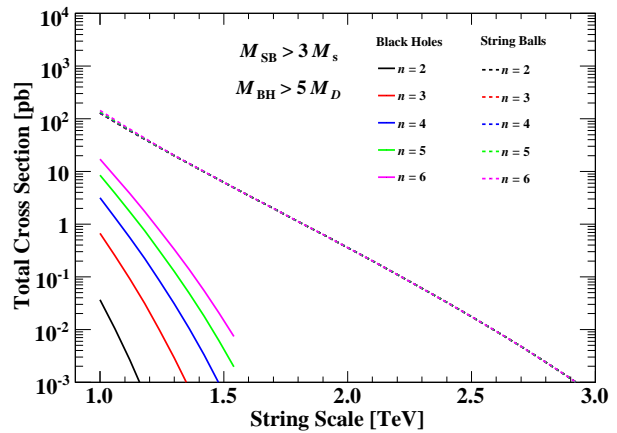}
\caption{Total proton-proton cross section versus string scale for the
production of black holes (solid curves) and string balls plus black
holes (dashed curves) for various numbers of extra dimensions $n$.
The black hole cross section increases with increasing $n$.
\label{figppMs}}
\end{center}
\end{figure}

Figure~\ref{figppMs} shows the total proton-proton cross section versus
string scale for the production of black holes and string balls for
various numbers of extra dimensions $n$. 
We see that the string ball plus black hole cross sections are about one 
to three orders of magnitude higher, and that a substantial range of
$M_\mathrm{s}$ could be probed with string balls.  
The black hole cross sections are strongly dependent on $n$ because of
the additional dependence on the relationship between the string scale
and the Planck scale Eq.~(\ref{eq11}).
The very weak dependence of the string ball cross section on $n$ is
mostly due to the $g_\mathrm{s}$ dependence of the string ball cross
section below the unitarity point, which depends on $n$ via
Eq.~(\ref{eq10}).  

\begin{figure}[htb]
\begin{center}
\includegraphics[width=\columnwidth]{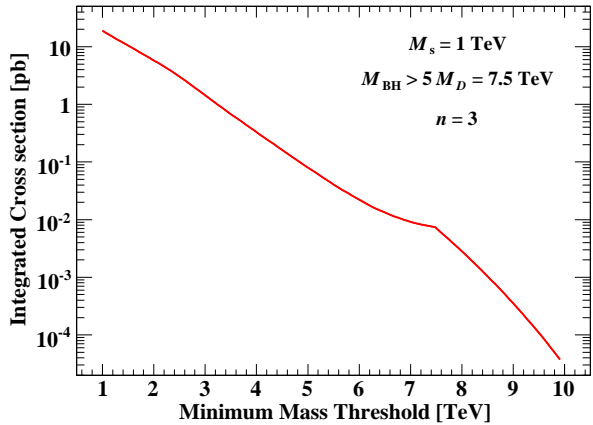}
\caption{Integrated proton-proton cross section versus minimum mass
threshold for the production of black holes and string balls plus black
holes for three extra dimensions.  
\label{figcum}}
\end{center}
\end{figure}

When searching for high-mass states above or near the Planck scale
experimentally, one is likely to search for an excess of events above a 
certain invariant mass threshold. 
Thus an important quantity is the integrated cross section above some
mass threshold. 
Figure~\ref{figcum} shows the integrated cross section versus minimum mass
threshold for $n=3$ extra dimensions.
Clearly visible is the correspondence point at 7.5~TeV, and not so
visible is the unitarity limit at 2.7~TeV.
Cross section values for masses less than about $3 M_\mathrm{s}$ may not
be reliable. 

Figure~\ref{figcum} shows that for $n=3$ and $M_D = 1.5$~TeV, the
integrated cross section for black hole production is about 10~fb.
Assuming a detector efficiency of 0.1, about 10~fb$^{-1}$ of data
might be required to discover or rule out GR black holes with these
parameters~\footnote{An acceptance of 0.1 is reasonable since early 
Monte Carlo estimates using the ATLAS detector indicate that an
acceptance of about 0.17 can be obtained by optimizing a set of cuts to
enhance the signal to background~\cite{CSC}.
To claim a discovery, we might require an excess of 10 events above
background and thus the required luminosity would be 10~fb$^{-1}$.
Correspondingly, if no events are observed above background and
systematic uncertainties are allowed for, about 10~fb$^{-1}$ of data
would be  required to rule out a 10~fb cross section at the 95\%
confidence level.}. 
Since the integrated cross section for string ball production at $3
M_\mathrm{s} = 3$~TeV is about 1~pb, the equivalent search for string
balls might require about 100 times less data.

%For this set of parameters, about 10~fb$^{-1}$ of data might be required
%to discover or rule out $n=3$ and $M_D = 1.5$~TeV GR black holes, while
%an equivalent search for string balls with $M_\mathrm{s} = 1$~TeV might
%require 100 times less data. 

To simulate string ball production and decay, we started from a modified
version of the Monte Carlo event generator CHARYBDIS version
1.003~\cite{Harris03a} and adapted it for our study.
The previous modifications were those described in
Ref.~\cite{Gingrich07a}, which included the addition of gravitons, black
hole recoil transverse to the 3-brane, and brane tension resisting the
black hole from leaving the brane. 
For the purposes of this study, the string cross section was added to
CHARYBDIS, along with the string parameters and the constraints between
them.
These additions were verified by reproducing all the cross section plots
in this paper. 
For the decays, the Hagedorn temperature for string ball decay replaced
the Hawking temperature for black hole decay.
The grey-body factors have been used in the results presented here.
The number of particle degrees of freedom and the probability for the
emission of each degree of freedom was not changed from
Ref.~\cite{Harris03a}. 

In the previously modified version of CHARYBDIS, the brane tension was a
free parameter.
By using string theory, we now have a model in which to predict this
tension.
The 3-brane tension ($D$-brane tension) is given
by~\cite{Polchinski,Alwis} 

\begin{equation} \label{eq24}
T = \frac{M_\mathrm{s}^4}{(2\pi)^3 g_\mathrm{s}}\, .
\end{equation}

\noindent
For $M_\mathrm{s} = 1$~TeV, $M_D = 1.5$~TeV, and $g_\mathrm{s} = 0.37$, 
the dimensionless brane tension in units of $M_D$ is $2.2\times 10^{-3}$.
For these parameters, there is about a 2\% probability for the black
hole or string ball to leave the brane according to the model in
Ref.~\cite{Gingrich07a}.
To avoid confusion between string ball effects and recoil effects, we
do not simulate graviton emission or black hole recoil in the remainder
of our studies. 

\begin{figure}[htb]
\begin{center}
\includegraphics[width=\columnwidth]{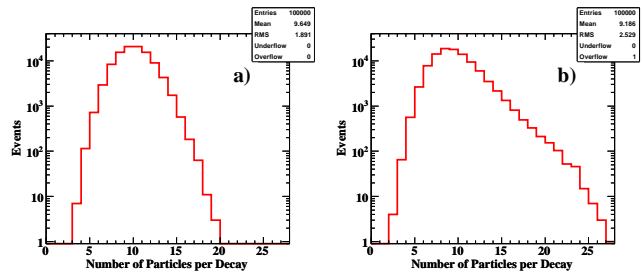}
\caption{Multiplicity distribution of visible primary particles emitted
from a) black holes with $7.5 < M < 14$~TeV and b) string balls and
black holes with $3 < M < 14$~TeV, for $n = 3$, $M_\mathrm{s} = 1$~TeV,
$M_D = 1.5$~TeV, and $g_\mathrm{s} = 0.37$.
\label{figmult}}
\end{center}
\end{figure}

In the following set of figures we will compare the experimentally
observable characteristics of black holes and string balls.
Figure~\ref{figmult} shows the multiplicity distributions for black
holes and string balls for $n=3$ extra dimensions.
Counted in the multiplicity are only primary visible particles (no neutrinos) 
evaporated from the black hole or string ball.
Decays of the primary particles are not counted.
The mean multiplicities are similar but the most probable number of
particles evaporated from the string ball is about one particle less.
This is probably due to the higher temperature of the string ball.
The string ball is capable of producing a maximum multiplicity of about
seven particles higher than black holes.
This is an infrequent occurrence but probably represents initial black
hole production followed by decay to a string ball.
An experimental requirement of a large number of objects in an event for 
black hole searches should also be a good requirement for string ball
searches. 

\begin{figure}[htb]
\begin{center}
\includegraphics[width=\columnwidth]{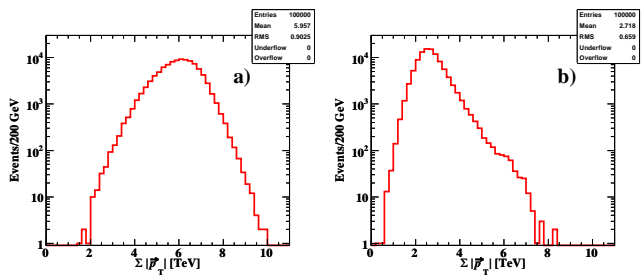}
\caption{Scalar sum of transverse momentum of all the visible particles
from a) black holes with $7.5 < M < 14$~TeV and b) string balls and
black holes with $3 < M < 14$~TeV, for $n = 3$, $M_\mathrm{s} = 1$~TeV,
$M_D = 1.5$~TeV, and $g_\mathrm{s} = 0.37$.
\label{figsum}}
\end{center}
\end{figure}

Figure~\ref{figsum} shows the distribution of the scalar sum of the
transverse momentum of each primary particle $i$, $\sum_i
|\vec{p}_\mathrm{T}|_i$, for black hole and string ball events.  
The mean drops from about 6~TeV for black hole events to about 2.7~TeV
for sting ball events.
%In searches for string ball events, it will be much more difficult to
%reduce backgrounds due to top-quark events and QCD events by restricting
%$\sum_i|\vec{p}_\mathrm{T}|$.

\begin{figure}[htb]
\begin{center}
\includegraphics[width=\columnwidth]{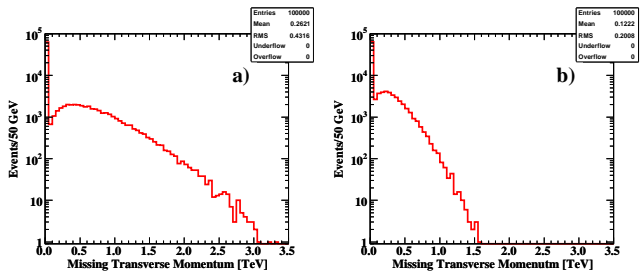}
\caption{Missing transverse momentum in a) black holes events with $7.5
< M < 14$~TeV and b) string balls and black holes events with $3 < M < 
14$~TeV, for $n = 3$, $M_\mathrm{s} = 1$~TeV, $M_D = 1.5$~TeV, and
$g_\mathrm{s} = 0.37$.
\label{figmiss}}
\end{center}
\end{figure}

Figure~\ref{figmiss} shows the missing transverse momentum distribution
in black hole and string ball events.
In this parton-level study, missing energy is due to electron, muon, and
tau neutrinos, and their antineutrinos.
The missing transverse momentum in string ball events is about one half
of that in black hole events, and is comparable to that resulting in
some supersymmetric models. 

\begin{figure}[htb]
\begin{center}
\includegraphics[width=\columnwidth]{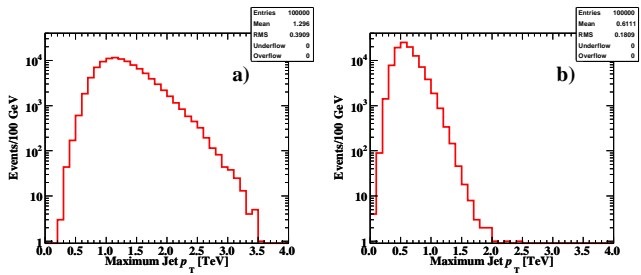}
\caption{Distribution of the highest transverse momentum jet from
the decays of a) black holes with $7.5 < M < 14$~TeV and b) string balls
and black holes with $3 < M < 14$~TeV, for $n = 3$, $M_\mathrm{s} =
1$~TeV, $M_D = 1.5$~TeV,  and $g_\mathrm{s} = 0.37$.
\label{figptj}}
\end{center}
\end{figure}

Figure~\ref{figptj} shows the transverse momentum $p_\mathrm{T}$ of the
highest transverse momentum jet in each event for black hole and string
ball decays. 
In this parton-level study, a jet is defined to be a quark, antiquark,
or gluon. 
The average highest-$p_\mathrm{T}$ jet in string ball events is about
$600-700$~GeV lower than in black hole events.
The maximum jet $p_\mathrm{T}$ in string ball events is about one half
of that in black hole events.
%It will thus be more difficult to reduce backgrounds using
%high-$p_\mathrm{T}$ jet requirements in string ball searches.  

\begin{figure}[htb]
\begin{center}
\includegraphics[width=\columnwidth]{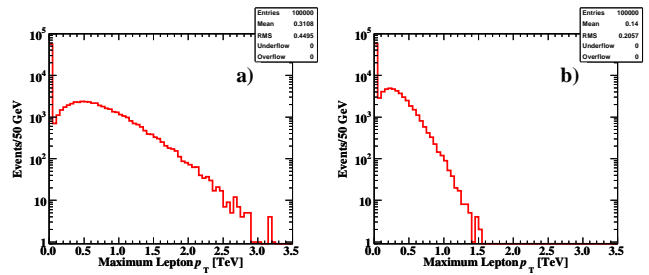}
\caption{Distribution of the highest transverse momentum lepton from
the decays of a) black holes with $7.5 < M < 14$~TeV and b) string balls
and black holes with $3 < M < 14$~TeV, for $n = 3$, $M_\mathrm{s} =
1$~TeV, $M_D = 1.5$~TeV,  and $g_\mathrm{s} = 0.37$.
\label{figptl}}
\end{center}
\end{figure}

Figure~\ref{figptl} shows the transverse momentum $p_\mathrm{T}$ of the
highest transverse momentum lepton in each event for black hole and string
ball decays. 
In this parton-level study, a lepton is defined to be an electron, muon, or
their antiparticles.
The average highest-$p_\mathrm{T}$ lepton in string ball decays is about
$100-200$~GeV lower than in black hole decays.
The maximum lepton $p_\mathrm{T}$ in string ball decays is about one
half of that found in black hole decays.
%It will thus be more difficult to reduce backgrounds using a
%high-$p_\mathrm{T}$ lepton requirement in string ball searches.  

To discuss the Standard Model backgrounds to string ball production, we 
rely on Ref.~\cite{CSC}.
Studying the backgrounds using a particle-level simulation would have
limited validity.
Potential Standard Model backgrounds to string ball events are processes
with large cross sections and multiple jets in the event, such as
top-quark production, QCD multijet production, as well as Z and W
production in combination with multiple jets. 

Since string balls (or black holes) are produced in the s-channel and
have high invariant mass, $\sum_i|\vec{p}_\mathrm{T}|_i$ is also large.
Requiring $\sum_i|\vec{p}_\mathrm{T}|_i > 2-3$~TeV significantly reduces
all but the QCD background and leaves a negligible top-quark background.   
The QCD background is significantly reduced but, because of its high
cross section, a small residual background remains.
Figure~\ref{figsum} shows that in searches for string ball events it
will be much more difficult to reduce backgrounds due to top-quark
events and QCD events by restricting $\sum_i|\vec{p}_\mathrm{T}|$, while
maintaining a reasonable acceptance for string ball events. 

To reduce the background further one can require multiple
high-$p_\mathrm{T}$ jets in the event. 
Figure~\ref{figptj} shows that it will be more difficult to reduce
backgrounds using high-$p_\mathrm{T}$ jet requirements in string ball
searches.    
While a requirement of $p_\mathrm{T} > 200$~GeV would be useful in black
hole events~\cite{CSC}, such a requirement applied to string ball events
would be detrimental to the search.

As a final requirement to further reduce the QCD background, a
high-$p_\mathrm{T}$ lepton requirement can be imposed. 
The probability of producing a high-$p_\mathrm{T}$ lepton in QCD events
is small. 
Requiring a lepton (electron or muon) with $p_\mathrm{T} > 50-100$~GeV
reduced the QCD background to a negligible level and reduced the
black hole single by less than one half.
Figure~\ref{figptl} shows that it will be more difficult to maintain a
high efficiency, while reducing backgrounds using a high-$p_\mathrm{T}$
lepton requirement in string ball searches.

%%%%%%%%%%%%%%%%%%%%%%%%%%%%%%%%%%%%%%%%%%%%%%%%%%%%%%%%%%%%%%%%%%%%%%%%%%%%%%%
\section{Discussion\label{sec7}}

To search for general relativistic black holes at the LHC one will have
to look above a certain minimum invariant mass threshold.
This threshold can become high relative to typical parton-parton centre of
mass energies at the LHC if the fundamental Planck scale is a few times
higher than its current lower bound.
If one tries lowering the threshold, the general relativistic
description of black holes breaks down and a quantum gravity description
is needed. 

One candidate for quantum gravity is weakly-coupled string theory.
By embedding string theory in large extra dimensions it is possible to
predict the occurrence of highly excited string states above the
fundamental string scale.  
According to the correspondence principle these string states match
those of a black hole at the energy in which a black hole can no longer
be considered as a general relativistic object.
Thus string theory allows us to lower the minimum invariant mass
threshold and search for string balls with properties not unlike those of
general relativistic black holes.
Furthermore, these string balls are the very objects in which the
general relativistic black holes will turn into as they evaporate and
make a transition across the mass threshold.

Lowering the minimum invariant mass threshold will enhance the search
range for trans-Planckian objects near the fundamental Planck scale.
However, the consequences of this will be a severely enhanced background
from Standard Model physics processes and supersymmetry, if discovered at a
TeV scale.
Typical search signatures for general relativistic black holes will
involve 
1) multiple high-$p_\mathrm{T}$ objects (jets, leptons, and photons), 
2) high energy in the events, measurable by $\sum_i
|\vec{p}_\mathrm{T}|_i$, 
3) large missing energy, 
4) spherical event shapes, etc.~\cite{CSC}.
While all of these quantities are significantly higher in general
relativistic black hole events than in typical Standard Model events,
this will not be true for string balls, unless the string scale is
significantly 
higher than about 1~TeV, but not too high as to exclude them from being
produced at the LHC.

Some might consider the paradigm of large extra dimensions to be
unlikely, and the possibility of producing black holes, let along string
states, to be highly unlikely.
However, those same people should admit that current experimental bounds
on fundamental parameters and searches have not yet ruled out their
possibility at the LHC.
On the other hand, black holes offer some of the most
interesting and broadest manifestations of fundamental physical
principles.
There is no reason to believe that highly-excited string states would
not also offer just as rich physics.
It is for these reasons that searches for general relativistic black
holes and other trans-Planckian phenomena must be initially taken seriously.
This paper provides a framework to allow experimentalists to develop
search strategies for highly-excited string states within the context of
weakly-coupled string theory embedded in large extra dimensions.

%%%%%%%%%%%%%%%%%%%%%%%%%%%%%%%%%%%%%%%%%%%%%%%%%%%%%%%%%%%%%%%%%%%%%%%%%%%%%%%
\begin{acknowledgments}
This work was supported in part by the Natural Sciences and Engineering
Research Council of Canada.
\end{acknowledgments}

%%%%%%%%%%%%%%%%%%%%%%%%%%%%%%%%%%%%%%%%%%%%%%%%%%%%%%%%%%%%%%%%%%%%%%%%%%%%%%%
\bibliography{gingrich}

%%%%%%%%%%%%%%%%%%%%%%%%%%%%%%%%%%%%%%%%%%%%%%%%%%%%%%%%%%%%%%%%%%%%%%%%%%%%%%%
\end{document}